# Topological classification for intersection singularities of exceptional surfaces in pseudo-Hermitian systems

Hongwei Jia[1,2,#], Ruo-Yang Zhang[1,#], Jing Hu[1], Yixin Xiao[1], Yifei Zhu[3,*], C. T. Chan[1,2,†]

[1]Department of Physics, the Hong Kong University of Science and Technology, Clear Water Bay, Kowloon, Hong Kong, China

[2]Institute for Advanced Study, the Hong Kong University of Science and Technology, Clear Water Bay, Kowloon, Hong Kong, China

[3]Department of Mathematics, Southern University of Science and Technology, Shenzhen, Guangdong, China

#These authors contributed equally to this work

*zhuyf@sustech.edu.cn; †phchan@ust.hk

Abstract: Exceptional points play a pivotal role in the topology of non-Hermitian systems, and significant advances have been made in classifying exceptional points and exploring the associated phenomena. Exceptional surfaces, which are hypersurfaces of exceptional degeneracies in parameter space, can support hypersurface singularities, such as cusps, intersections and swallowtail catastrophes. Here we topologically classify the intersection singularity of exceptional surfaces for a generic pseudo-Hermitian system with parity-time symmetry. By constructing the quotient space under equivalence relations of eigenstates, we reveal that the topology of such gapless structures can be described by a non-Abelian free group on three generators. Importantly, the classification predicts a new kind of non-Hermitian gapless topological phase and can systematically explain how the exceptional surfaces and their intersections evolve under perturbations with symmetries preserved. Our work opens a new pathway for designing systems with robust topological phases, and provides inspiration for applications such as sensing and lasing which can utilize the special properties inherent in exceptional surfaces and intersections.

*Introduction*. Singularities, characterized by the property of non-differentiability, are ubiquitous and play important roles in many physical systems in the real world. The occurrence of a singularity is usually accompanied by exotic physical phenomena [1-13]. In topological materials, a Weyl point in a Hermitian system acts as a sink or source of the Berry curvature, and two Weyl points with opposite chiralities are connected by a Fermi-arc surface state [1,2,9,11]. Topology is pivotal in understanding the existence and stability of singularities, and a singularity can be characterized by a topological invariant (e.g. Chern number), which is usually encoded in the adiabatic evolution of eigenstates on closed loops/surfaces enclosing it [5,6-9,11]. Recently, the topology of non-Hermitian systems is attracting growing attention [14-25]. Non-Hermiticity is ubiquitous because most systems are not isolated where the eigen-energies become complex numbers, representing energy exchange with surrounding environment. The exceptional points, as a unique feature of non-Hermiticity, are singular points on the complex energy plane where both the eigenenergies and the eigenstates coalesce [14-19]. Different from Hermitian degeneracies such as Weyl points and nodal lines, the exceptional point may carry fractional topological invariants [16,18,19,24,26], and can induce stable bulk Fermi-arcs [22,24] and braiding of eigenvalues [26]. The non-Hermitian skin effect, manifested by the strong dependence of the eigen-spectrum on boundary conditions, is associated with the point gaps in bulk topology [15-18,21,23,25]. Recent discoveries of lines, rings and surfaces of exceptional points further enriched the classes of topological degeneracies [27-31]. High-order exceptional degeneracies, which often appear as cusps of exceptional lines/surfaces, carry hybrid topological invariants in a higher dimensional parameter space [32].

Lots of efforts have been devoted to classifying exceptional points recently. Topological classifications are important, because whenever the type of energy gaps and Altland-Zirnbauer symmetry class for the system are known, the degeneracies in the parameter space are predictable [14,19,20,33-35]. It is thus a theoretical framework for predicting non-Hermitian topological phases of matter, providing guidance for experimental realizations. Particularly, exceptional points can assemble into hypersurfaces in 3D parameter space, dubbed exceptional surfaces (ESs), separating exact and broken phases [20]. The ES are commonly observed in non-Hermitian systems with parity-time inversion symmetry (*PT*) or chiral symmetry [20,27-29], and have broad applications in the design of sensing and absorption devices [31,36]. As a subspace of the parameter space, ESs may possess embedded lower-dimensional singularities, the so-called hypersurface singularities, that have remarkable properties differentiating them from other points on the ESs, such as intersections [37], cusps [38-40], and swallowtail catastrophes [41]. Such hypersurface singularities are symmetry protected [31,37-41], and are stable against perturbations with symmetries preserved. However, despite various important physical phenomena and potential applications, the hypersurface singularities on ESs were never topologically classified.

The non-defective intersection line (NIL) of ESs should be the simplest form of hypersurface singularities, which is an elementary singular line composing the swallowtail catastrophe [41]. In this work, we consider a generic two-level non-Hermitian system, which preserves *PT* symmetry plus an additional pseudo-Hermitian symmetry. The band structures of such systems exhibit non-defective intersection lines (NIL) of ESs. By analyzing equivalence relations of eigenstates, we discover that the quotient space of the order-parameter space is homotopy equivalent to a bouquet of three circles $M = S^1 \vee S^1 \vee S^1$. The topology of the NIL is thus represented by the fundamental group of *M*, which is a non-Abelian free group on three generators. The topological charges represented by the group elements can be associated with frame deformation of eigenstates. Besides, the conservation of topological charges is manifested by the robustness of ESs and NILs against perturbations to the Hamiltonian.

*Main.* The prototypical Hamiltonian is a two-level system, which is *PT* symmetric and also preserves an additional $\eta$-pseudo-Hermitian symmetry [41-43]

$$[H, PT] = 0, \qquad \eta H \eta^{-1} = H^\dagger \tag{1}$$

The *PT* operator can be regarded as complex conjugation with a proper choice of basis in parameter space, and thus the Hamiltonian can always be gauged to be real. The metric operator $\eta$ here takes the Minkowski metric $\eta = \text{diag}(-1,1)$ [13,41,44,45]. It is notable that in case the Hamiltonian performs unitary transformations, $\eta$ transforms simultaneously (see Section 1 in [46] for details). These symmetries imply that the momentum space Hamiltonian can be written in the form

$$H(\mathbf{k}) = f_2(\mathbf{k})i\sigma_2 + f_3(\mathbf{k})\sigma_3 \tag{2}$$

where $f_{2,3}$ are real functions in three-dimensional (3D) **k**-space, and $\sigma_{2,3}$ are Pauli matrices. The term multiplied by an identity matrix can be ignored because it has no impact on the gapless structure. Such Hamiltonians correspond to physical systems with non-reciprocal hopping [41,47-49] of orbitals.

The 2D $f_2$-$f_3$ plane is exactly the order parameter space of all the Hamiltonians preserving the symmetries in Eq. (1) due to the following reasons: Any ESs in 3D **k**-space can be mapped to the exceptional lines (ELs) at $f_2 = \pm f_3$, and these surface intersections (NILs) correspond to the intersecting point (NIP) of the two ELs at the origin $f_2 = f_3 = 0$; a path in **k**-space can be mapped to a path on the $f_2$-$f_3$ plane, and if the path circulates around the NIL, the corresponding path on $f_2$-$f_3$ plane circulates around the NIP. The gapless structure of the order parameter space is shown in Fig. 1a, with the red and green lines denoting EL$_1$ and EL$_2$ satisfying $f_2 = \mp f_3$, respectively. In regions I and III (satisfying $|f_2| < |f_3|$), the eigenenergies are real, which are *PT*-exact phases. In contrast, regions II and IV (satisfying $|f_2| > |f_3|$) are *PT*-broken phases, with eigenvalues forming complex-conjugate pairs. The paths $\alpha$, $\alpha'$, $\beta$ and $\beta'$ are terminated at different ELs in parameter space. The NIP at the origin is what we want to classify, and is excluded from the 2D plane [20,50]. Any paths or loops cannot traverse the NIP. First, the 2D plane with a hole at the origin can deformation retract to a circle $S^1$ [Fig. 1(b)]. Such a mathematical process can be interpreted as a quotient map, which identifies all points on the ray starting from the origin (note that the origin is excluded). As a consequence, all points on upper half of EL$_1$ shrink to point $A$, and the lower half shrinks to point $A'$. Similar process applies to EL$_2$, with the upper and lower halves shrinking to $B$ and $B'$. There also exist equivalence relations on the $S^1$. At point $A$, the two eigenstates coalesce, which is totally the same as the coalesced state at point $A'$. Therefore, $A$ and $A'$ can be identified, and we glue $A'$ to $A$ via a quotient map. The same procedure applies to $B$ and $B'$. It is notable that antipodal points lying in the gap cannot be identified. The essential difference is that each of these points has two linearly independent eigenstates, and the two eigenstates are ordered by the corresponding eigenenergies. The sequence of eigenenergies in exact phases (Regions I and III) is well-defined. In broken phases, the real parts of eigenenergies coalesce, and we simply order the eigenstates by the imaginary parts of eigenenergies. One can either sort the imaginary parts from negative to positive or from positive to negative, and the only requirement is that the criterion in Regions II and IV of broken phases should be unified. Consequently, even though a pair of antipodal points in the gap have the same eigenstates, but the eigenstates are ordered inversely for the two points based on the eigenenergies, which thus cannot be identified. With the above procedures, we obtain the quotient space (details of quotient space are shown in Section 2 of [46]) of the $S^1$ in Fig. 1(b), which is a bouquet of three circles [see Fig. 1(c)],

$$M = S^1 \vee S^1 \vee S^1 \tag{3}$$

The fundamental group of $M$ can thus be derived as

$$\pi_1(M) = \mathbb{Z} * \mathbb{Z} * \mathbb{Z} \tag{4}$$

which is a free non-Abelian group on three generators. As indicated by Fig. 1(c), the generators $Z_1$, $Z_2$ and $Z_3$ can be characterized by the paths combinations $\alpha\beta$, $\alpha\alpha'^{-1}$ and $\alpha'\beta'$, respectively. These topological invariants can be assocated with the frame deformations of eigenstates, details are shown in [46].

We next introduce some typical trivial and nontrivial loops (or combined paths) in parameter space to better understand the group [Eq. (4)]. The combined paths characterizing the generators $Z_1$, $Z_2$ and $Z_3$ are shown in Fig. 2(a-c), respectively, where the dashed lines with arrows denote quotient maps that glue identified points. We note that the gluing process does not mean the loop traverses the NIP. Each of the combined paths corresponds to an $S^1$ in Fig. 1c, which are exactly closed loops in quotient space $M$. A closed loop encircling the NIP is also a path combination $\alpha\beta\alpha'\beta'$, which characterizes the topological invaraint $Z_1Z_3$, being an element in the group [Eq. (4)]. Topologically protected edge states can arise from $Z_1Z_3$, and details are shown in Section 4 of [46]. Some other nontrivial loops are discussed in Section 5 of [46]. It is found that all nontrivial loops (paths) have to traverse ELs. Any loop that does not touch ELs is confined in a specific region, e.g. loop $l$ in Fig. 3e, and is always trivial because it cannot enclose any singularity. As we move the loop across the EL, $l$ is decomposed into two paths $l_1$ and $l_2$ (Fig. 3f). As the terminal points of $l_1$ (or $l_2$) can be identified, $l_1$ (or $l_2$) is also a loop in quotient space $M$, but is a trivial loop that can shrink to a point. Therefore, the combination $l_1l_2$ is also trivial. By stretching the loop to cross the other EL (see Fig. 3g), $l$ becomes a composite of $l_1l_3l_4l_5$. Since both $l_1$ and $l_4$ correspond trivial loops in quotient space $M$, the composite is equivalent to the combination $l_3l_5$. In addition, paths $l_3$ and $l_5$ are along opposite directions and are homotopic to $\alpha^{-1}$ and $\alpha$, respectively. It is thus not difficult to find out that the combination $l_1l_3l_4l_5$ remains trivial. *From this analysis, we can conclude that a loop (or a path) encountering ELs through continuous deformations will not change the topology. In contrast, encountering NIPs is not allowed, which will change the topology.* Similar discussion can also be found in [41]. These results indicate that the NIP, as a hypersurface singularity, is distinguished from isolated singularities. An open path for an isolated singularity is meaningless because it cannot enclose the singularity. Conversely, a path joining ELs (or ESs) can provide a lot of information on the NIP (see more details in Section 3 of [46]). Therefore, if a loop is partitioned into several segments by ELs (or ESs), it is necessary to investigate the evolution of eigenstates on each path and then discuss the combinations. Since the non-trivial loops all traverse the hypersurfaces, our method is affiliated to intersection homotopy theory [50,51].

With these elementary invariants taken from the group, we are able to predict how singularities evolve under the constraint of topological invariants as the Hamiltonian deforms. We are more interested in the 3D **k**-space, in which the NILs and ESs correspond to NIP and ELs on the $f_2$-$f_3$ plane, respectively. Consider the following example

$$f_2(\mathbf{k}) = k_x k_z, \quad f_3(\mathbf{k}) = -k_x^2 + k_y^2 + k_z^2 - d \quad , \tag{5}$$

for which the chain of NILs and ESs in 3D **k**-space for $d=1$ are shown in Fig. 3a1, where the red and green surfaces denoted by ES$_1$ and ES$_2$ satisfying $f_2 = \mp f_3$ and corresponds to EL$_1$ and EL$_2$ (Fig. 1), respectively. We first look at the loop $l_6$ in Fig. 3a1, which encloses the waists of two ESs and their intersections NILs (satisfy $f_2 = f_3 = 0$), but does not touch any ESs. Such a loop is trivial according to our previous conclusion, which is not obvious in the figure because these ESs and NILs seemingly prevent the loop from retracting to a point. However, by changing $d$ from positive to negative, the waists of ESs and NIL gradually retract to a point (Fig. 3b1), and finally open up to form a bandgap (Fig. 3c1). The two NILs enclosed by the loop are thus annihilable. This is a solid evidence that $l_6$ is a trivial loop. The trivial loop $l_6$ enforces the ESs connecting the two NILs to remain smooth as the Hamiltonian deforms. This can be explained by $l_6'$ (see Fig. 3a1), which is equivalent to $l_6$, as they enclose the same NILs, but traverses ESs. On the section plane that $l_6'$ lying on, $l_6'$ is segmented by ESs into several paths, as sketched in Fig. 3a2, where the red and green lines denote the ESs cut by the plane. The invariants on the segments of $l_6'$ must cancel each other to form a trivial product, implying that the path $l_t$ is trivial, which can be directly seen from the fact that the path $l_t$ is a trivial path with both ends on a smooth ES (see Fig. 3a2). As one continues to deform the Hamiltonian, the two ESs enclosed become disconnected after the two NILs annihilate (see Fig. 3c1-c2). Now we turn to $l_7$ in Fig. 3a1, which is segmented by the ESs into different paths, as indicated by Fig. 3a3. The loop is a path combination $(\beta^{-1}\alpha^{-1}\beta'^{-1}\alpha'^{-1})^2$, carrying a squared topological invariant $(Z_1^{-1}Z_3^{-1})^2$. This nontrivial invariant forbids the two encircled NILs to annihilate as $d$ varies in the Hamiltonian (Eq. 5). The two NILs merge at a point for $d=0$ (Fig. 3b1), and the nearby area is partitioned into eight regions (see Fig. 3b3), meaning that the point still carries the squared invariant $(Z_1^{-1}Z_3^{-1})^2$. As one keeps varying $d$, the point splits, and the two NILs become separate in the other direction, as shown in Fig. 3c1 and c3. The squared invariant $(Z_1^{-1}Z_3^{-1})^2$ is conserved within the deformation process of the Hamiltonian. *We thus understand that the squared invariant $(Z_1^{-1}Z_3^{-1})^2$ is a necessary condition for the chain of NILs.* Another condition for the presence of the chain of NILs is the mirror symmetries $k_x \rightarrow -k_x$ and $k_z \rightarrow -k_z$, which will be illustrated in a tight binding model to be described below. The conservation of the invariant also shows that two inannihilable NILs cannot be directly connected by smooth ESs, as can be observed in Fig. 3a3-c3.

*Non-reciprocal tight binding model.* The model Hamiltonian can be realized in real space using a tight binding model with non-reciprocal hoppings. A 3D fcc lattice model and the corresponding Brillouin zone are shown in Fig. 4a-b, where $M$ and $N$ denote two inequivalent lattice sites with opposite onsite energies $\pm E_0$. The hopping between $M$ and $N$ (on dark green bonds) is non-reciprocal ($M{\rightarrow}N$: $t_1$, $M{\rightarrow}N$: $-t_1$), and the hoppings on yellow and red bonds [between the adjacent sites in the same sublattice but in different directions, i.e. yellow bonds: $\vec{r}_M \rightarrow \vec{r}_M + \vec{a} + \vec{b}$ and $\vec{r}_N \rightarrow \vec{r}_N + \vec{a} - \vec{b}$; red bonds: $\vec{r}_M \rightarrow \vec{r}_M + \vec{a} - \vec{b}$ and $\vec{r}_N \rightarrow \vec{r}_N + \vec{a} + \vec{b}$] are characterized by $t_2$ and $-t_2$, respectively. The corresponding real space Hamiltonian is given by

$$H_1 = \sum_{\substack{\vec{r}_M \in \bar{G}_M \\ \bar{\alpha}=\bar{a},\bar{b},\bar{c}}} t_1(a^{\dagger}_{M,\vec{r}_M} a_{N,\vec{r}_M+\bar{\alpha}} + a^{\dagger}_{M,\vec{r}_M} a_{N,\vec{r}_M-\bar{\alpha}}) - h.c. + E_0(a^{\dagger}_{M,\vec{r}_M} a_{M,\vec{r}_M} - a^{\dagger}_{N,\vec{r}_N} a_{N,\vec{r}_N})$$
$$+ \sum_{\substack{\vec{r}_h \in \bar{G}_h \\ h=M,N}} \mathrm{sgn}(h) t_2(a^{\dagger}_{h,\vec{r}_h} a_{h,\vec{r}_h+\bar{a}+\bar{b}} + h.c. - a^{\dagger}_{h,\vec{r}_h} a_{h,\vec{r}_h+\bar{a}-\bar{b}} - h.c.)$$

(6)

where $\vec{a}$, $\vec{b}$ and $\vec{c}$ are the set of orthogonal lattice vectors connecting lattice sites $M$ and $N$ (see Fig. 4a). Here sgn($h$)=1 and $-1$ for $h=M$ and $N$, respectively. The corresponding **k**-space Hamiltonian (see details in Section 6 of [46]) shows that $f_3(\mathbf{k}) = E_0 + 2\sin k_x \sin k_y$ and $f_2(\mathbf{k}) = \cos k_x + \cos k_y + \cos k_z$. The system has mirror symmetries in the $x$ and $y$ directions for $E_0$=0, and thus the band structure is symmetric about $k_x$=$\pi/d_L$ and $k_x$=0 planes. The ESs and NILs for $E_0$=0 are plotted in Fig. 4c, where the red and green surfaces are ESs satisfying $f_2 = \mp f_3$, respectively As can be seen, a chain of NILs is formed on the intersection line of the $k_x$=$\pi/d_L$ and $k_y$=0 planes (see Fig. 4c). The orange dashed loop (Fig. 4d) is a combination $(\alpha'\beta'\alpha\beta)^2$ that carries a squared topological invariant $(Z_3Z_1)^2$. The blue dashed loop does not traverse any ES and is trivial. The mirror symmetries and the topological invariants on the two loops ensure the presence of intersection points of NILs (red arrows). A nonzero $E_0$ can break the mirror symmetries in $k_x$ and $k_y$ directions, which eliminates the intersection points (as shown in Fig. 4e). However, the breaking of mirror symmetries does not affect the topology on the loops. As shown in Fig. 4e, the blue loop is still trivial, because it does not touch any ESs. The topological invariant on the orange loop is conserved [still $(Z_3Z_1)^2$], as the traversed ESs remain the same (Fig. 4d and 4f). The variation of ESs and NILs with respect to perturbations to the Hamiltonian is thus predictable in view of the conservation of topological charges.

In summary, we topologically classified a generic non-Hermitian two-level system with *PT* and an additional pseudo-Hermitian symmetries, which can be realized in systems with non-reciprocal hoppings [41,47-49]. Such systems exhibit surfaces of exceptional points, which intersect stably in momentum space. We showed that the topology of such gapless structure can be understood from the quotient space under equivalence relations of eigenstates, which is a bouquet of three circles. Its fundamental group is isomorphic to a free non-Abelian group on three generators. The group structure allows the prediction of the evolution of ESs and NILs as the Hamiltonian deforms. The method of quotient space topology might potentially be extended to classify other hypersurface singularities in non-Hermitian systems, such as high-order exceptional points as cusps [32,40] and more complicated swallowtail catastrophes [41]. Our work predicts a new kind of non-Hermitian gapless topological phase of matter, providing pathways for designing systems to realize robust topological non-defective degeneracies in non-Hermitian systems. Future theoretical and experimental works on investigating the bulk-edge correspondence for hypersurface singularities may stem from this work. The classification also provides guidance for the design of relevant devices in sensing and lasing applications using exceptional surfaces and their intersections.

Acknowledgements: This work is supported by Research Grants Council of Hong Kong through grants AoE/P-502/20, 16307621, 16307821, 16307420, 16310420 and Croucher Foundation (CAS20SC01) and KAUST20SC01. Y. Zhu acknowledges the financial support from National Natural Science Foundation of China (NSFC) grant 11701263.

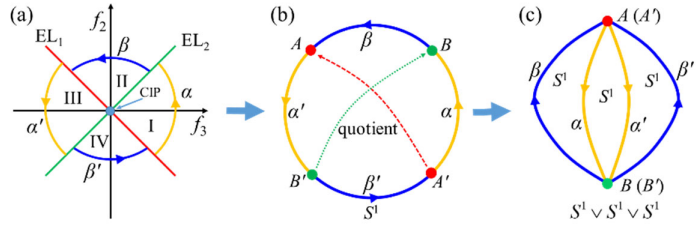

Fig. 1. Construction of quotient space under equivalence relations. (a) The gapless structure of parameter space, EL$_{1,2}$ characterize exceptional lines satisfying $f_2 = \mp f_3$, respectively. The NIP is at the origin. Regions I and III are *PT*-exact phases, and Regions II and IV are *PT*-broken phases. The ELs intersect at $f_2=f_3=0$. (b) The 2D plane excluding the NIP can deformation retract to a circle $S^1$, with the upper and lower parts of EL$_1$ shrinking to $A$ and $A'$, respectively. Similarly upper and lower parts of EL$_2$ shrink to $B$ and $B'$, respectively. (c) Gluing identified points $A$ with $A'$, and $B$ with $B'$, the quotient space of $S^1$ in (b) can be obtained as a bouquet of three circles.

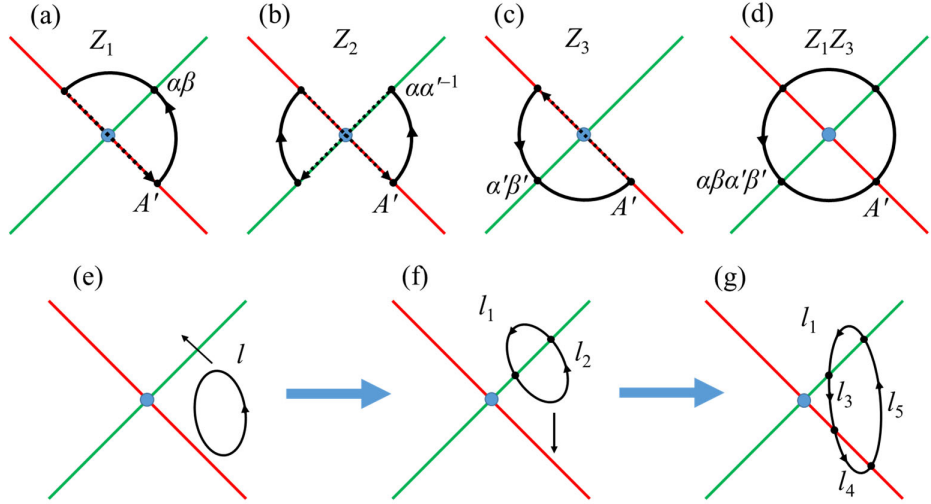

Fig. 2. Typical nontrivial and trivial loops. (a-c) Nontrivial loops characterizing generators $Z_1$, $Z_2$ and $Z_3$, where the dashed lines with arrows denote quotient maps, i.e. gluing identified points. (d) The loop formed by the combination $\alpha\beta\alpha'\beta'$ encloses the NIP, which characterizes the topological invariant $Z_1Z_3$. Point $A'$ in (a-d) denotes the basepoint. (e) A loop without touching ELs is confined within a specific region and is trivial. (f) Moving the loop along the black arrow direction [see (e)], the loop becomes a composite of paths $l_1$ and $l_2$. Both of $l_1$ and $l_2$ are trivial, and their composite loop is also trivial. (g) Stretching the loop along the black arrow direction in (f), the loop crosses $EL_1$ and becomes a composite of paths $l_1l_3l_4l_5$. The path $l_4$ corresponds to a trivial loop in quotient space $M$. The paths $l_5$ and $l_3$ are in opposite directions, and are homotopic to $\alpha$ and $\alpha^{-1}$, respectively. The path product $l_1l_3l_4l_5$ is thus trivial.

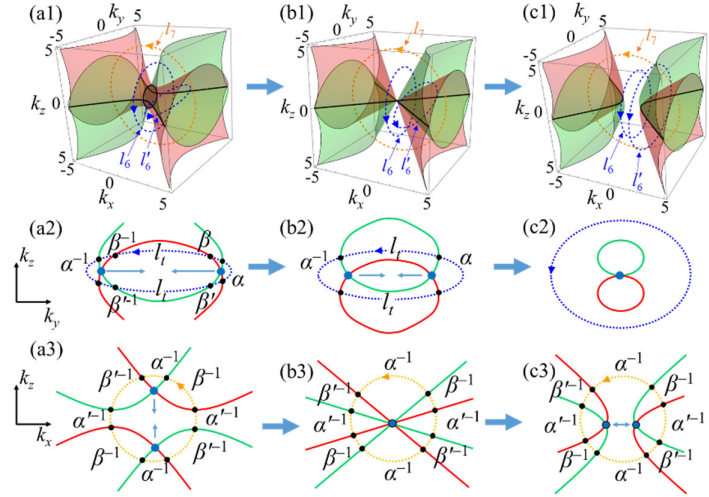

Fig. 3. Evolution of ESs and NILs against perturbations to the Hamiltonian. (a1-c1) ESs (red and green surfaces) and NILs (black lines) plotted with Eq. 5, corresponding to $d>0$, $d=0$ and $d<0$, respectively. The blue loops $l_6$ and $l_6'$ have trivial topological invariants. (a2-c2) Cross section of the plane that $l_6'$ locates on. The enclosed pair of NILs can annihilate each other. (a3-c3) Cross section of the plane that the orange $l_7$ locates on. The NILs enclosed cannot annihilate each other. Red and green lines: ESs; Dark blue dots: NILs; Black dots: intersecting points of loops on ESs (in Row 2 and Row 3).

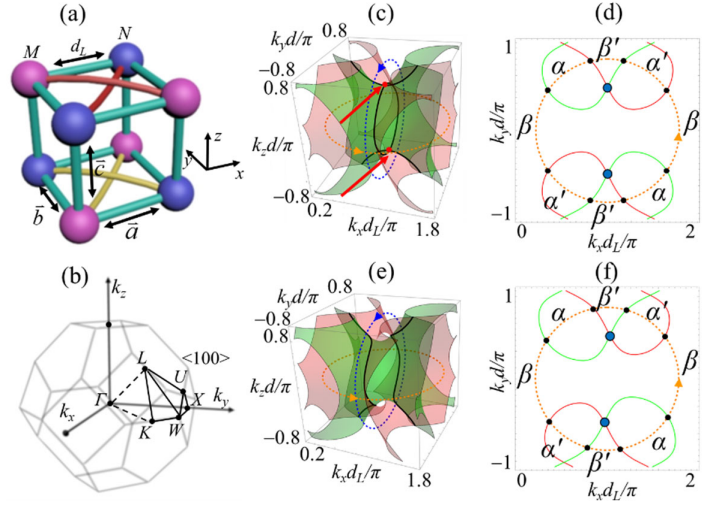

Fig. 4. Proposal of an fcc lattice model with non-reciprocal hoppings. (a) fcc lattice with two sites $M$ (blue balls) and $N$ (pink balls). The interspace distance between $M$ and $N$ is $d_L$, and $\vec{a}$, $\vec{b}$ and $\vec{c}$ are bond vectors. The hopping on dark green bonds is non-reciprocal ($M \to N$: $t_1$, $N \to M$: $-t_1$). The hopping on the same lattice sites in different directions (in $\vec{a}+\vec{b}$ and $\vec{a}-\vec{b}$) have opposite signs (hopping on yellow bonds: $t_2$, hopping on red bonds: $-t_2$). (b) First Brillouin zone of the fcc lattice. (c, e) ESs (red and green surfaces) and NILs (black lines) for $E_0=0$ and $E_0 \neq 0$ in Eq. (6). (c) has a chain of NILs, which is symmetric with respect to <100> plane. The intersecting points on the chain are labelled with red arrows. (d, f) Cross section of the plane $k_z=0$ (where the orange loop locates) for (c) and (d), respectively. The topological charge on the loop is conserved even though the mirror symmetries are broken.

# Supplemental material: Topological classification for intersection singularities of exceptional surfaces in pseudo-Hermitian systems

Hongwei Jia[#], Ruo-Yang Zhang[#], Jing Hu, Yixin Xiao, Yifei Zhu[*], C. T. Chan[†]

## 1. Pseudo-Hermiticity and metric operator

The pseudo-Hermiticity can be regarded as a symmetry in non-Hermitian physics [1], and a formal definition of pseudo-Hermiticity is always accompanied with a metric operator $\eta$

$$\eta H \eta^{-1} = H^{\dagger} \tag{S1}$$

Hence, a pseudo-Hermitian system is also called a $\eta$-pseudo-Hermitian system, and the metric operator $\eta$ is a Hermitian matrix. Recently, the parity-time inversion symmetry ($PT$) is included in pseudo-Hermiticity symmetry [2,3]. The considered system thus includes two inequivalent pseudo-Hermitian symmetries. In quantum mechanics, the Hamiltonians of two systems can be considered to be equivalent if they can transform to each other via unitary transformations ($U^{-1} = U^{\dagger}$)

$$H\varphi = E\varphi \rightarrow UHU^{\dagger}U\varphi = EU\varphi \rightarrow H'\varphi' = E\varphi' \tag{S2}$$

We apply the transformation to Eq. S1

$$\begin{aligned} U\eta H \eta^{-1} U^{\dagger} &= UH^{\dagger}U^{\dagger} \\ \rightarrow U\eta U^{\dagger} UHU^{\dagger} U\eta^{-1} U^{\dagger} &= UH^{\dagger}U^{\dagger} \\ \rightarrow \eta' H' \eta'^{-1} &= H'^{\dagger} \end{aligned} \tag{S3}$$

where $\eta' = U\eta U^{\dagger}$ is the transformed metric operator. For the considered system in Eq. (S2), one can apply an SU(2) transformation to the Hamiltonian, e.g.

$$\begin{aligned} H' &= e^{i\frac{\theta}{2}\sigma_1} H e^{-i\frac{\theta}{2}\sigma_1} \\ &= (f_2(\mathbf{k})i\sigma_2 + f_3(\mathbf{k})\sigma_3)\cos\theta + (-f_2(\mathbf{k})i\sigma_3 + f_3(\mathbf{k})\sigma_2)\sin\theta \end{aligned} \tag{S4}$$

It is found that the Hamiltonian can be transformed to a $PT$-symmetric system with equal gain and loss for $\theta = \pi/2$,

$$H' = -f_2(\mathbf{k})i\sigma_3 + f_3(\mathbf{k})\sigma_2 \tag{S5}$$

and the metric operator is simultaneously transformed to

$$\eta' = \begin{bmatrix} 0 & i \\ -i & 0 \end{bmatrix} \tag{S6}$$

Hence, the classification in this work can be extended to other $PT$-symmetric pseudo-Hermitian systems (e.g. realized by equal gain and loss, Eq. S5) with equivalent metric operators [4].

## 2. Quotient space and stratified space

In topology, the quotient space of a topological space under given equivalence relations is a new topological space constructed by endowing the quotient set of the original topological space with the quotient topology [5]. Let $(X, \tau_X)$ be a topological space, and let ~ be equivalent relation on $X$. The quotient set $Y=X/\sim$ is the set of equivalence classes of elements of $X$. The equivalence class of $x \in X$ is denoted by $[x]$. The quotient map associated with ~ refers to the surjective map

$$q: X \to X/\sim$$
$$x \to [x]$$

(S7)

Intuitively speaking, all points in each equivalence class are identified or glued together. A well-known example of quotient space is the Brillouin zone. In the momentum space of periodic systems, a point **k** is identified with points $\mathbf{k}+m_a\mathbf{G}_a$ because a **k**-space Hamiltonian at these points have the same eigenvalues and eigenstates. Here $\mathbf{G}_a$ are reciprocal lattice vectors and $m_a$ are integers. That is why we mostly considers the band dispersions in the first Brillouin zone. It is also notable that the points on one side of the Brillouin zone boundaries can be translated to the points on the other boundary under translational operations of $\mathbf{G}_a$. Such points are identified and can be glued together. As simple examples, the first Brillouin zone is a quotient map of the momentum space under equivalence relation of translations by $\mathbf{G}_a$, and points in the first Brillouin zone are the representatives of all the equivalence classes. For 1D periodic systems, identifying points on the first Brillouin zone boundary constructs a quotient space, which is a 1D circle $S^1$ (see Fig. S1a1-a2). Similarly, opposite edges ($p_1$ and $p_2$, and $p_3$ and $p_4$) of the Brillouin zone of 2D periodic systems can be identified (see Fig. S1b1). By gluing $p_1$ to $p_2$, the Brillouin zone becomes a cylinder (see Fig. S1b2). We further glue $p_3$ to $p_4$, and the cylinder becomes a torus $T$ (see Fig. S1b3). $p_1$ (or $p_2$) and $p_3$ (or $p_4$) are called the skeleton of the torus, and is a bouquet of two circles with a common basepoint $S^1 \vee S^1$. The surface of the torus is called the two-cell. Assembling the skeleton and the two-cell, the torus can be described by the product $T = S^1 \times S^1$. The topology of the torus is thus described by its fundamental group $\pi_1(T) = \mathbb{Z} \times \mathbb{Z}$. This is a free Abelian group on two generators.

The momentum space of the considered system is a stratified space [6,7]. In topology, a stratified space is a triple $(V,S,\zeta)$, where $V$ is a topological space (often we require it to be locally compact, Hausdorff, and second countable), $S$ is a decomposition of $V$ into strata $V = \bigcup_{X \in S} X$, and $\zeta$ is the set of control data $\{(T_X),(\pi_X),(\rho_X)|X \in S\}$, where $T_X$ is an open neighborhood of the stratum $X$, $\pi_X$: $T_X \to X$ is a continuous retraction, and $\rho_X$: $T_X \to [0,+\infty)$ is a continuous function. These data need to satisfy the following conditions:

1. Each stratum $X$ is a locally closed subset and the decomposition $S$ is locally finite.

2. The decomposition $S$ satisfies the axiom of the frontier: if $X, Y \in S$ and $Y \cap \bar{X} \neq \emptyset$, then $Y \subset \bar{X}$. The condition implies that there is a partial order among strata: $Y<X$ if and only if $Y \subset \bar{X}$ and $Y \neq X$.

3. Each $T_X$ is a smooth manifold.

4. $X=\{v \in T_X| \rho_X(v)=0\}$. So $\rho_X$ can be viewed as the distance function from the stratum $X$.

5. For each pair of strata $Y<X$, the restriction $(\pi_X, \rho_X)$: $T_Y \cap X \to Y \times (0,+\infty)$ is a submersion.

6. For each pair of strata $Y<X$, there holds $\pi_Y \circ \pi_X = \pi_Y$ and $\rho_Y \circ \pi_X = \rho_Y$.

Consider the parameter space $f_2$-$f_3$ of our Hamiltonian, the topological space $V$ is simply the plane (Fig. S2). Thus $S$ is the decomposition of $V$ into three strata ($X, Y, Z$), which are the 2D space $\mathbb{R}^2$ ($X$), the singular hypersurfaces ELs at $f_2 = \pm f_3$ ($Y$=Sing($X$)), and the hypersurface singularity NIP ($Z$=Sing(Sing($X$))) at the center, as shown in Fig. S2. For each stratum (e.g. $X$), the smooth manifold $T_X$ considers the nearby neighborhood. Therefore, $T_1$-$T_3$ in Fig. S2 correspond to the three strata $X$, $Y$ and $Z$, respectively.

Our classification is based on eigenstates. The Hamiltonian in spaces without gap closing can be expressed with the sum

$$H = \sum_{i=1,2} E_i \left| \varphi_i^L \right\rangle \left\langle \varphi_i^R \right| \tag{S8}$$

where $\varphi_i^{L(R)}$ denote the left and right eigenstates of the Hamiltonian. The pseudo-Hermiticity and $PT$ symmetries of the system enforces the left and right eigenstates (both in exact and broken phases) to be connected by the following relation

$$\varphi_i^L = \eta (\varphi_i^R)^* \tag{S9}$$

The quotient space is constructed by identifying points with the same eigenstates. Note that the eigenstates are ordered by the corresponding eigenvalues, and the criterion for ordering eigenstates has been introduced in the maintext. Hence, gluing point $A'$ and point $A$, and $B$ to $B'$ is understandable, because the two eigenstates at these points coalesce, and ordering eigenstates is meaningless at these points.

$$\begin{aligned} \varphi_1 = \varphi_2 = \begin{bmatrix} -1 \\ 1 \end{bmatrix} & \quad \text{for } f_2 = f_3 \\ \varphi_1 = \varphi_2 = \begin{bmatrix} 1 \\ 1 \end{bmatrix} & \quad \text{for } f_2 = -f_3 \end{aligned} \tag{S10}$$

However, in spaces without gap closing, by adding a minus sign to the Hamiltonian in Eq. S8, both eigenenergies take negative signs, and the eigenstates remain the same. This process can be realized by taking the negatives of $f_2$ and $f_3$, which are just the antipodal points lying in opposite regions with respect to the NIP. Even though the two points have the same eigenstates, the order of the two states exchanges for antipodal points because eigenvalues are added by minus signs. Therefore, the two points cannot be identified, which is distinct from the points on ELs. The constructed space Eq. (3) in the main text is a stratified quotient space, and the corresponding topology Eq. (4) is thus a quotient space topology. Since the nontrivial loops in parameter (or quotient) space all traverses the singular hypersurfaces (i.e. EL or ES), our approach is affiliated to the intersection homotopy theory [6].

### 3. Frame deformation of eigenstates

The metric operator for pseudo-Hermiticity plays a similar role as the space-time metric in general relativity [8,9], and the eigenstates are like local coordinate frames (or tetrad). The local metric $g$ can be defined with the indefinite inner product $g_{mn} = \langle \varphi_m | \eta \varphi_n \rangle$ [10]. The symmetries provide an orthogonality relation to the right eigenstates

$$\varphi_m^T \eta \varphi_n \begin{cases} = 0 & m \neq n \\ \neq 0 & m = n \end{cases} \tag{S11}$$

Since the Hamiltonian is gauged to be real, the eigenstates are real by removing arbitrary phases in exact phases ($|f_3|>|f_2|$), and thus the eigenstates in $PT$-exact phases have another orthogonal relation

$$\langle \varphi_m | \eta \varphi_n \rangle \begin{cases} = 0 & m \neq n \\ \neq 0 & m = n \end{cases} \quad (S12)$$

meaning that $g$ is a diagonal matrix. However, the two diagonal elements of $g$ have opposite signs, i.e. one vector is space-like and the other is time-like, which imposes the Riemannian geometry [i.e. SO(1,1) transformation] of the evolution of eigenstates as parameters vary. It is notable that in a specific region, $g$ is invariant. In our previous work discussing the topology of swallowtail catastrophes in non-Hermitian systems [10], we established the relationship between the local metric $g$ and the geometric phase. Here we will not repeat the derivation details, and we simply use the relationship between the metric $g$ and the affine connection of the geometric phase

$$A^{*\,l}_{k_i\,m} g_{ln} + g_{ml} A^{\,l}_{k_i\,n} = 0 \quad (S13)$$

to predict the emergence of ELs and NIPs. The affine connection is defined by the eigenstates

$$A^{\,n}_{k_i\,m} = -\langle \varphi_n | \eta \frac{\partial}{\partial k_i} | \varphi_m \rangle \quad (S14)$$

More details can be found in [10].

In *PT*-exact phases, the local metrics in Region I and Region III are in the following forms

$$g_\mathrm{I} = \begin{bmatrix} 1 & 0 \\ 0 & -1 \end{bmatrix}, \quad g_\mathrm{III} = \begin{bmatrix} -1 & 0 \\ 0 & 1 \end{bmatrix} \quad (S15)$$

Here the sequence of eigenvalues is defined by sorting the corresponding eigenvalues from small to large. The geometric phase is an integration of the affine connection

$$U^{-1} = \mathrm{P} \exp(\int_{\mathbf{k}(0)}^{\mathbf{k}(\xi)} d\mathbf{k} A_\mathbf{k}) \quad (S16)$$

where P is the path ordering operator because the affine connection is a matrix. It is not difficult to find out that the two eigenstates experience Lorentz boost and the geometric phase is simply

$$U^{-1} = \exp \gamma T \quad (S17)$$

where $T$ is the Lie algebraic generator of SO(1,1) group

$$T = \begin{bmatrix} 0 & 1 \\ 1 & 0 \end{bmatrix} \quad (S18)$$

and can be derived from Eq. (S13-14). Next, we define the path parameter $\theta$ (see Fig. S3a), with $f_3 = \cos\theta$ and $f_2 = \sin\theta$. The evolution of eigenvalues and eigenstates along the path $\alpha$ ($-\pi/4 \leq \theta \leq \pi/4$) is shown in Fig. S3b1 and b2, respectively. Note that the eigenstates have been rescaled. As can be indicated in Fig. S3b2, the two eigenstates are rotating in opposite directions, and resultantly, they evolve from parallel states to antiparallel states, which is typical for frame deformations. This process occurs because $\gamma$ varies from $+\infty$ to 0 and to $-\infty$, and the infinity of $\gamma$ is provided by the ELs, i.e. the path departs from EL$_1$ and terminates at EL$_2$. It is thus understandable that the frame deformation is a result of hyperbolic transformation, i.e. the Lorentz boost in general relativity [8]. In Region III, the evolution of eigenstates is similar to that in Region I, simply the two eigenstates swap.

In broken phases, the local metrics are both

$$g_{II,IV} = \begin{bmatrix} 0 & 1 \\ 1 & 0 \end{bmatrix}, \tag{S19}$$

and the evolution of eigenstates is still defined on SO(1,1) group. The difference is that the two eigenstates become complex conjugate, and the frame deformation process is extended to the complex space. Results for path $\beta$ ($\pi/4 \leq \theta \leq 3\pi/4$) is provided in Fig. S3c. As shown in Fig. S3c2-c3, the initially parallel eigenstates bifurcate to form a conjugate pair, and finally evolve to two anti-parallel imaginary vectors.

With the above frame deformation process on any of the paths aforementioned, one can already determine that an NIP can be formed by the intersection of the two ELs (or ESs). Hence, an open path joining ELs (or ESs) can provide a lot of information on the intersection of the ELs (or ESs). This is essentially different from isolated singularities, for which a path is only meaningful whenever it is closed. Our former work [10] has established the relation between the frame deformation with the conventional Berry phase, and in the next section we will discuss this accompanied with topologically protected edge states.

## 4. Topologically protected edge states of NIP

The isolated singularities in band structures (e.g. Dirac/Weyl points or nodal lines) are often accompanied with edge states, which are stable against boundary conditions of the system, and are protected by the non-trivial topology of the singularities. However, the bulk-edge correspondence of hypersurface singularities in non-Hermitian systems has not been discussed. Apparently, it seems impossible to find stable edge states for such singularities, because closed loops circulating them will inevitably traverse ESs and experience real line gap closing. As a matter of fact, hypersurface singularities indeed support topologically protected edge states. In this section, we will provide solid evidence for this counter-intuitive physical phenomenon based on the NIP (or NIL).

We consider the following 1D k-space Hamiltonian corresponding to a lattice model,

$$H(k) = \sigma_z \cos k + i\sigma_y \sin k + v\sigma_0 \cos(k+a) \tag{S20}$$

where $\sigma_0$ is the 2 by 2 identity matrix, and the last term proportional to $\sigma_0$ is useful in tuning bandgaps so that edge states can be easily found out. As a common fact, introducing the identity term does not change the topology of the system, and the degeneracy features remain unchanged. It is obvious that $f_2(k)=\cos k$ and $f_2(k)=\sin k$, and the Hamiltonian preserves the symmetries in Eq. (1) of the maintext. The trajectory of $(f_2(k), f_3(k))$ encloses the NIP, as shown in Fig. S4a. Such a Hamiltonian can be realized by the 1D periodic system in Fig. S4b, and the corresponding real space Hamiltonian is

$$H_r = \underbrace{\frac{1}{2}(\sigma_z + \sigma_y + ve^{ia}\sigma_0)}_{t_1} \sum_j c_j^\dagger c_{j+1} + \underbrace{\frac{1}{2}(\sigma_z - \sigma_y + ve^{-ia}\sigma_0)}_{t_2} \sum_j c_j^\dagger c_{j-1} \tag{S21}$$

where $j$ denotes unit cell index, and $t_1$ and $t_2$ are both 2 by 2 hopping matrices, with their elements denote hopping parameters between lattice sites (see Fig. S4b). It is shown that the hopping matrices have the following relation.

$$t_1^* = t_2 \tag{S22}$$

Since the loop inevitably traverses the ELs four times, the band structure on the 1D Brillouin zone experience real line gap closing four times, as indicated by Fig. S4c. For the path $\alpha$ residing in Region

I, our former result shows that the two initially parallel eigenstates bifurcate and evolve to two anti-parallel states (Fig. S3b2), which is a frame deformation process. This process shows that the relative rotation angle between the two eigenstates is equal to π, which is equal to an integral

$$\psi = \oiint_{l_\alpha} i\langle\varphi|\nabla_k\varphi\rangle d\mathbf{k} \tag{S23}$$

The loop $l_\alpha$ in the integration is shown in Fig. S4d, which is simply joining the trajectories of the two eigenvalues on path $\alpha$ at ELs. Therefore, the loop $l_\alpha$ is in the 3D Re($\omega$)-$f_2$-$f_3$ space, not simply confined in the $f_2$-$f_3$ plane. It is not difficult to identify that Eq. (S23) is the conventional Berry phase. Thus the frame deformation is related with the conventional Berry phase. Related discussions have already been provided in [10]. On path $\alpha'$, the two eigenstates swap compared with path $\alpha$, and thus the relative rotation angle is $-\pi$, meaning that the Berry phase along the loop $l_{\alpha'}$ is $-\pi$ (see Fig. S4d). In addition, the identity term in the Hamiltonian (Eq. S20) introduces a real line gap between the eigenenergies on path $\alpha$ and path $\alpha'$. Resultantly, if we truncate the 1D system with open boundaries, there will be a pair of edge modes (the truncated system has two edges) residing in the line gap, as shown in Fig. S4e. Here OBC stands for open boundary condition, and PBC stands for periodic boundary condition. Since the eigenenergies in broken phases have point gaps, skin effect also arises, as indicated in Fig. S4e. The edge states are away from any bulk modes and skin modes, and thus skin modes and edge modes can be easily distinguished. The field distribution (amplitude $|\varphi|$) of one edge mode is shown in Fig. S4f, where $Ns$ denotes lattice sites. To obtain the eigenenergies and eigenstates under OBC, we simply write down the truncated real space Hamiltonian with finite unit cells (here we preserve 300 unit cells), and numerically solve the eigenvalue problem of the real space Hamiltonian.

## 5. Some other nontrivial loops in parameter space

In Fig. 2 of the maintext, we introduced some typical nontrivial loops and the corresponding topological invariants. Since the number of elements in the group (Eq. 4) is infinitely large, and some elements other than Fig. 2 is also useful, here we give a brief introduction on these invariants and the corresponding path combinations in parameter space.

Figure S5(a) shows the path product $\alpha'\beta$. Note that the basepoint has been fixed at $A$ (or $A'$), and thus we cannot exchange the order in the product (i.e. $\beta\alpha'$). Exchanging the order in the product means that the basepoint is changed from $A$ (or $A'$) to $B$ (or $B'$). In homotopy theory, one will obtain another fundamental group by changing the basepoint without changing the order parameter space, and the groups obtained by changing the basepoint are isomorphic to each other since the quotient space $M$ is path-connected. It is not difficult to find out that $\alpha'\beta=\alpha'\alpha^{-1}\alpha\beta$, and thus the corresponding topological invariant is $Z_2^{-1}Z_1$, which is an element of the group (Eq. 4). The path combination $\beta'^{-1}\beta$ is totally in broken phases, and is a counterpart of Fig. 2(b). Since $\beta'^{-1}\beta$ can be obtained as the path product $\beta'^{-1}\alpha'^{-1}\alpha'\alpha^{-1}\alpha\beta$, it is thus obtained that the invariant on the loop is $Z_3^{-1}Z_2Z_1$. In a similar way, the path combination in Fig. S5(c) $\alpha\beta'$ can be obtained as the product $\alpha\alpha'^{-1}\alpha'\beta'$, and the invariant on the loop is simply $Z_2^{-1}Z_3$.

## 6. k-space Hamiltonian matrix of the lattice model

In the maintext, we provide a lattice model to show that such Hamiltonians can be realized by non-reciprocal hoppings. We also illustrated that the chain of CILs can be formed because it is protected by the squared invariants $(Z_1Z_3)^2$ [or $(Z_1^{-1}Z_3^{-1})^2$] and the trivial invariant (on the blue loop) (see Fig. 4c in the maintext). The intersection chain (denoted by the black arrows in Fig. 4c) points are also protected by the mirror symmetries $k_x+\pi/d\rightarrow-k_x+\pi/d$ and $k_y\rightarrow-k_y$ for $E_0=0$. Breaking the mirror symmetries

(setting $E_0$ to be nonzero) will open the chain points but will not affect the topology on the blue and orange loops. Here we provide the **k**-space Hamiltonian of the lattice model. Firstly, the real space Hamiltonian can be transformed to a **k**-space Hamiltonian by performing Fourier transformation (setting the length of bond vectors $d=1$),

$$\begin{aligned}H_1 = & t_1(e^{ik_x} + e^{-ik_x} + e^{ik_y} + e^{-ik_y} + e^{ik_z} + e^{-ik_z})a^\dagger_{M,k}a_{N,k} - h.c. \\ & + E_0(a^\dagger_{M,k}a_{M,k} - a^\dagger_{N,k}a_{N,k}) \\ & + t_2(e^{ik_x+ik_y} + e^{-ik_x-ik_y} - e^{ik_x-ik_y} - e^{-ik_x+ik_y})(a^\dagger_{M,k}a_{M,k} - a^\dagger_{N,k}a_{N,k})\end{aligned} \quad (S24)$$

By diagonalizing Eq. (S24), we obtain

$$H_1(\mathbf{k}) = \begin{bmatrix} E_0 + 2\sin k_x \sin k_y & \cos k_x + \cos k_y + \cos k_z \\ -\cos k_x - \cos k_y - \cos k_z & -E_0 - 2\sin k_x \sin k_y \end{bmatrix} \quad (S25)$$

Hence, it can be expanded with Pauli matrices $i\sigma_2$ and $\sigma_3$ [with $f_3(\mathbf{k}) = E_0 + 2\sin k_x \sin k_y$ and $f_2(\mathbf{k}) = \cos k_x + \cos k_y + \cos k_z$, see Eq. (2) in the maintext], preserving the symmetries in Eq. (1).

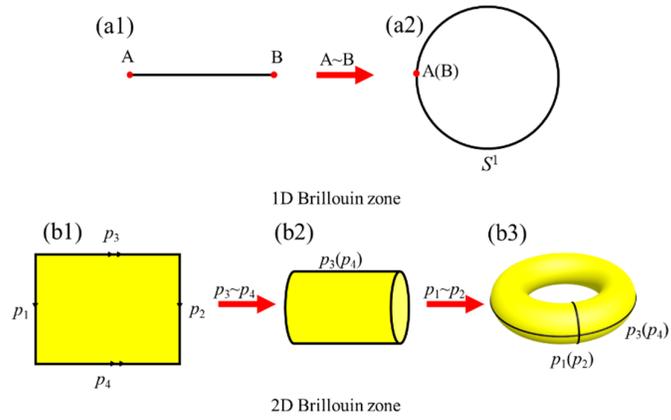

Fig. S1. Quotient space of momentum space in periodic systems. (a1)-(a2) The quotient space of 1D Brillouin zone is a circle ($S^1$) by identifying the two points on the Brillouin zone boundary. (b1)-(b3) Construction of quotient space of 2D Brillouin zone. Identifying the boundaries $p_3$ with $p_4$ gives a cylinder, which becomes a torus by identifying $p_1$ with $p_2$.

$V := \mathbb{R}^2$

$S := \{\mathbb{R}^2, > \{|f_2| = |f_3|\}, > \{f_2 = f_3 = 0\}\}$

$T_1 := \mathbb{R}^2$ 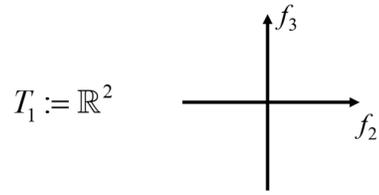

$T_2 :=$ 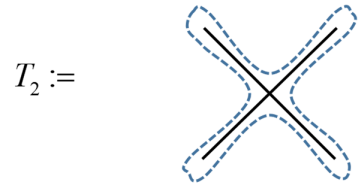

$T_3 :=$ 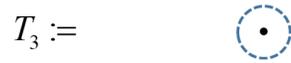

Fig. S2. Stratified space of the 2D plane with ELs and NIP.

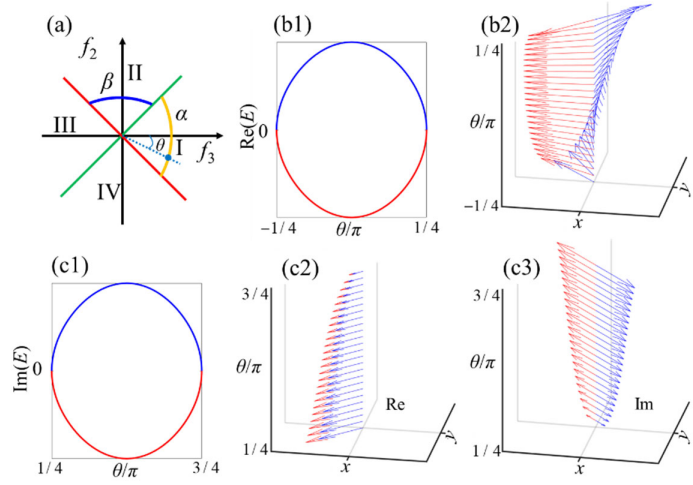

Fig. S3. Frame deformation along different paths. (a) Paths $\alpha$ and $\beta$ in parameter space. $\theta$ denotes the path parameter, i.e. $f_3=\cos\theta$, $f_2=\sin\theta$, $-\pi/4 \leq \theta \leq \pi/4$ for $\alpha$, $\pi/4 \leq \theta \leq 3\pi/4$ for $\beta$. (b1-b2) Evolution of eigenvalues (real part, b1) and eigenstates along path $\alpha$. (c1-c3) Evolution of eigenvalues (imaginary part, c1) and eigenstates (c2, real part; c3, imaginary part) along path $\beta$.

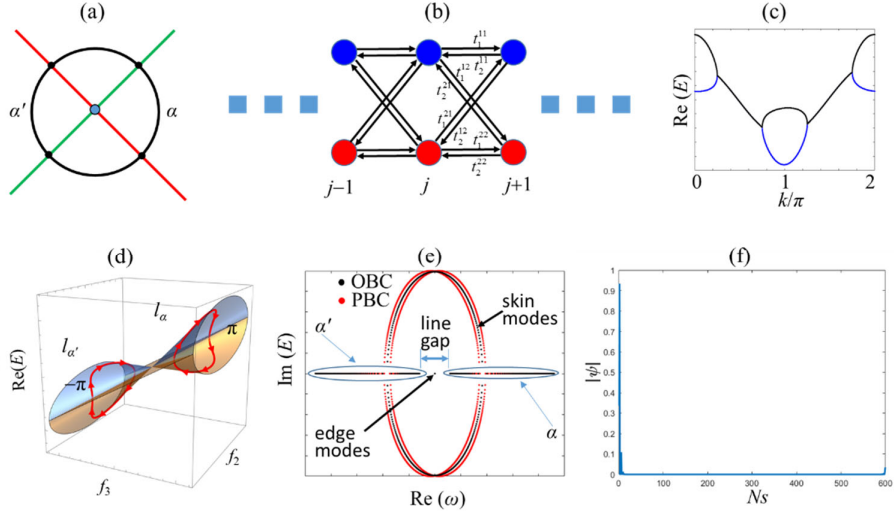

Fig. S4. Topologically protected edge states of NIP. (a) A loop circulating the NIP, being the Brillouin zone of the 1D lattice model in Eq. S12, is partitioned into four paths, with $\alpha$ and $\alpha'$ residing in exact phases. (b) Realization of the lattice model. (c) Eigenvalue dispersions (real part) of the model Eq. S20. (d) Joining the trajectories of two bands on path $\alpha$ forms a loop in Re($\omega$)-$f_2$-$f_3$ space $l_\alpha$, along which the Berry phase is $\pi$. For path $\alpha'$, joining the two bands forms the loop $l_{\alpha'}$, along which the Berry phase is $-\pi$. (e) Eigenvalues of the model with open boundary conditions (OBC) and periodic boundary conditions (PBC). There exist a pair of edge modes in the line gap for eigenstates on paths $\alpha$ and $\alpha'$. (f) Field distribution of one edge mode. The lattice model with OBC has 300 periods (600 lattice sites).

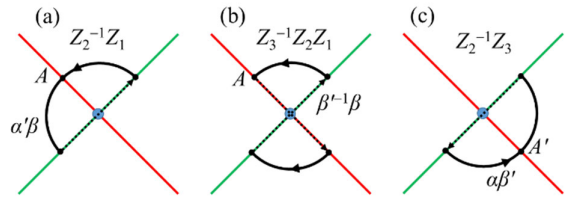

Fig. S5. Some other nontrivial loops and the corresponding topological invariants (other than Fig. 2) taking from the group Eq. 4.